# Allocating the chains of consecutive additions for optimal fixed-point data path synthesis

Ilya Y. Zhbannikov and Gregory W. Donohoe
Department of Electrical Engineering, University of Idaho, Moscow, ID, 83843
Email: zhba3458@vandals.uidaho.edu,  gdonohoe@uidaho.edu

*Abstract—* **Minimization of computational errors in the fixed-point data path is often difficult task. Many signal processing algorithms use chains of consecutive additions.  The analyzing technique that can be applied to fixed-point data path synthesis has been proposed. This technique takes advantage of allocating the chains of consecutive additions  in order to predict growing width of the data path and minimize the design complexity and computational errors.**

## I. Introduction

Real-time software applications in embedded systems such as mobile devices must be implementable in small, lightweight, low-power, low-cost platforms. Often these applications are computationally-greedy: real-time speech and video processing, embedded control. In desktop microprocessors, floating-point representation of real numbers is commonly used. For these reasons, embedded systems are often built with microprocessors, Digital-Signal Processors (DSP) or Field-Programmable Gate Arrays (FPGA), and hybrid system-on-a-chip (SOC) platforms such as the Xilinx Virtex series, which do not have hardware floating-point support.

An alternative to floating-point numbers implementation of algorithms is fixed-point arithmetic, which can be implemented with integer-only hardware. The benefits are high speed, low hardware complexity and low cost. When implemented well, fixed-point arithmetic can be accurate enough for most embedded applications.

Designing a fixed-point data path by hand is slow and error-prone, however. Software design tools can help, but the fixed point tools available are useful only for simulation, not for synthesis. During designing and synthesis the fixed-point data path, the designer usually reserves particular amount of bits for the whole part and also some amount of bits for the fractional part of the real number in order to represent a real value in fixed-point domain. Fixed-point notation should be able to represent any real number within an acceptable range and resolution, capture overflows and track dependencies between variables and scaling factors. Notation can have various implementations (particular implementation depends on designer's choice), for example the notation of $(1/I/F)$ , where $I$ – represents the number of bits in the whole part of the real value, $F$ – denotes the number of bits reserved for fractional part and leading 1 represents one bit reserved for sign (the notation $(I/F)$ for unsigned representation). In other words, the machine-word space has the following partitioning: 1IIIFFFF. For example the binary value 01011111 with notation of (1/3/4) represents the decimal value of +5.9375 and  10100001 represents the -5.9375.

*This work was fully supported by the Department of Electrical Engineering, University of Idaho.*

Another notation was developed in Sandia National Laboratory [2] and named as "SIF-notation", in this notation the value of $S$ is the number of bits denoted to sign and can be more or equal to one.

Usually, a signal processing algorithm consists of standard arithmetic operations, such as addition, multiplication, subtraction and division. The complexity of fixed-point data path design is due to the fact of that designer must take into accounts possible overflows, that happen during run-rime (because the input data usually cannot be predicted for all possible cases). As well as the need of minimization of computational error for maintaining the specification's accuracy of the algorithm. All these things make serious obstacles during designing and take significant design time. To simplify the design complexity the conception of allocating chains of consecutive additions was created. This allows prediction of growing the width of the result (measured in bits, i.e. how many bits are in the result of arithmetic operation) in a long chain of consecutive additions. In other words, there is the possibility to track the critical places in the fixed-point data path where overflows occur (and so the whole system fails or at least does not implement all specification's functions). Such prediction significantly decreases complexity and time during fixed-point data path design. Also it allows reducing the computational (rounding) errors.

In this work the conception of allocation the chains of additions in the data path is explored. In the next sections the whole conception will be explained and  experimental results will be discussed.

## II. Proposed conception

In worst-case scenario the bit-length of the result of binary addition increases due to the fact of carry-out. Consider an example of addition two signed 8-bit integers (that come from 8-bit ADC, with fixed-point notation of 1/7/0) which produces the result with length of 9 data bits as shown in Fig. 1.

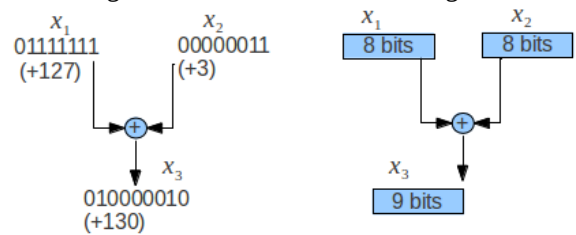

Figure. 1. Example of addition.

The binary value of 130 occupies 9 bits (eight bits in the whole part and one sign bit), which is one bit larger that the bit-lengths of the operands. In general case, when two operands are being added, the overflow exists when result does not fit the machine word space. Suppose the machine word is 16 bits long. Reserving 7 bits for data (see Fig. 2) and leaving other 9 bits empty (or as sign bits), a simple fixed-point notation will be: SSSSSSSSSIIIIIIII or (9/7), where S denotes a sign bit and I denotes a data bit in the whole part of the number.

After addition the machine-word partitioning will be SSSSSSSSIIIIIIII or (8/8) because carry-out occurred and one extra bit was added to the whole part. During designing a fixed-point data path, there is the overflow constraint which means that the designer always have to keep up with carry-outs and reserve an extra bit in order to prevent overflow. Otherwise design can fail during run-time.

Consider allocation the chain of consecutive additions as the way of fixed-point data path designing and minimizing possible rounding (truncation) errors. Many signal processing algorithms contain chains of consecutive additions (for example, the convolution algorithm). Assuming that the chain of consecutive additions is already given, refer to Fig. 3. Considering the chain of consecutive additions, one can assume that overflow may occur at every stage, i.e. worst-case scenario may always happen and one extra bit is being added to every addition stage.

Assumption of worst-case scenario at every step during the data path design may lead to unneeded truncations of viable data bits and bring extra computational error when result does not fit the available space in machine word. However, it turns out that overflow occurs only in *specific* locations of given chain, which is shown in Fig. 4.

In Fig. 4 the chain of additions with particular steps where overflows occur is shown. For simplification, the operands are unsigned 8-bit integers (no sign bit). In the eighth addition (in the end of the addition chain), the data path width is increased by 4 extra bits. Notice, that the width of the data path increases by one in steps of 1, 2, 4, 8. To build the theory and obtain a reasonable equation that allows us tracking the steps when overflow occurs, let us denote the number of additional bits as $k$. This is shown in Fig. 5. Denote the number of single step where overflow occurs as $S_n$. $N$ is the position of the highest bit (little-endian scheme) in the operand $x_2$, $N = 0...\infty$, in this case $N = 7$.

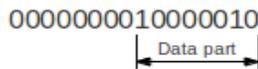

Figure 2. Picking a data part from machine word.

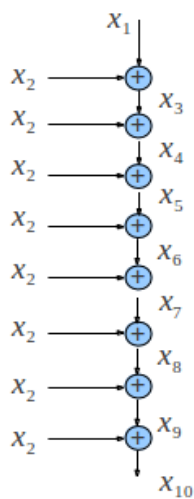

Figure 3. Chain of consecutive additions.

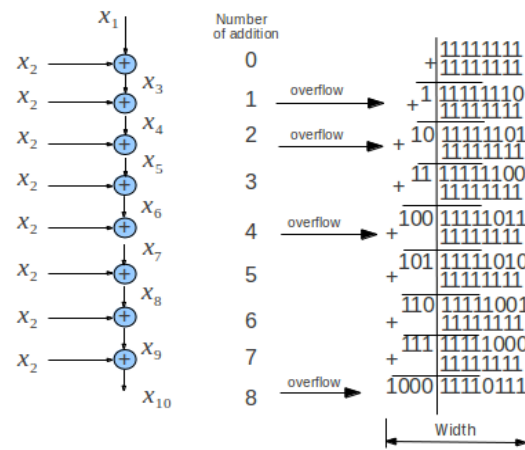

Figure 4. Chain of consecutive additions with overflows.

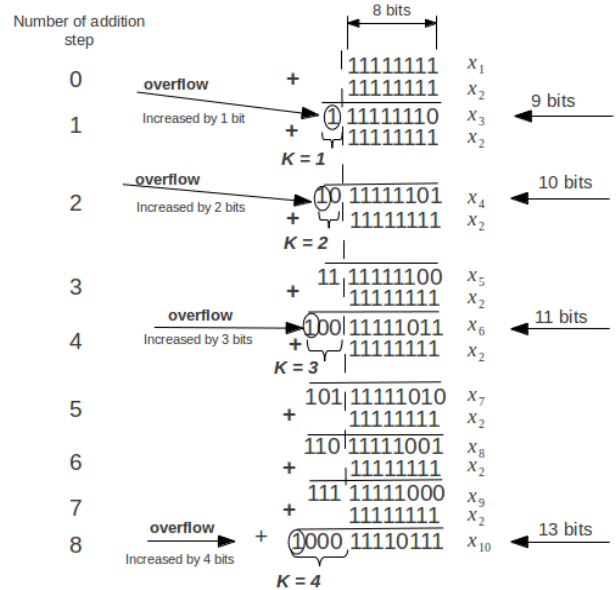

Figure 5. Chain of additions with tracking the bit-length of the result.

For *k*-th overflow the result would be:

$$Result_k = 2^0 + 2^1 + 2^2 + \ldots + 2^{N+k-1} \quad (1)$$

For the *k*+1 overflow, the result would be:

$$Result_{k+1} = 2^0 + 2^1 + 2^2 + \ldots + 2^{N+k} \quad (2)$$

The number of steps $S_{k,k+1}$ (number of consecutive additions) between two consecutive overflows $k$ and $k+1$ would be:

$$S_{k,k+1} = \frac{Result_{k+1} - Result_k}{2^0 + 2^1 + 2^2 + \ldots + 2^N} = \frac{2^{N+k}}{2^{N+1}-1} \quad (3)$$

The equation (3) should be modified:

$$S_{k,k+1} = \left[\frac{2^{N+k}}{2^{N+1}-1}\right]_{floor} \quad (4)$$

For example, for $k = 3$ and $k + 1 = 4$, the number of steps between them is 4.

This equation (4) can be extended for the whole addition chain. Denote $S_n$ as the number of particular step when overflow occurs (*n* means that overflow occurs on *n*-th step). For $k = 0$, the result would be:

$$Result_0 = 2^0 + 2^1 + 2^2 + \ldots + 2^N \quad (5)$$

For $k = n$, the result would be:

$$Result_n = 2^0 + 2^1 + 2^2 + \ldots + 2^{N+n-1} \quad (6)$$

Subtracting (5) from (6) and dividing the intermediate result by $2^{N+1}-1$, the number of steps between $k = 0$ and $k = n$ (which is the number of *n*-th overflow in the chain of consecutive additions) will be:

$$S_n = \left[\frac{2^{N+n-1} + 2^{N+n-2} + 2^{N+n-3} + \ldots + 2^0 - 2^N - 2^{N-1} - \ldots 2^0}{2^{N+1}-1}\right]_{floor} \text{ or}$$

$$S_n = \left[\frac{2^{N+n}}{2^{N+1}-1}\right]_{floor} \quad (7)$$

As an example, the fourth overflow will occur at:

$$S_4 = \left[\frac{2^{7+4}}{2^{7+1}-1}\right]_{floor} = 8 \text{ -th step.}$$

In other words, designer can *predict* overflows during the analyzing the whole data flow graph by simply allocating existing chains of consecutive additions. Consider an example of 5-tap FIR filter (Fig. 6). Machine word length is 16 bits. The fixed-point notation for input nodes ($x_n, x_{n-1}, x_{n-2}, x_{n-3}, x_{n-4}$) is (9/0/7) and for weights $w_0, w_1, w_2, w_3, w_4$ is (8/0/8). The fixed-point format of nodes 1, 2, 3, 4, 5 is (1/015) since the binary length of the result in binary multiplication is the sum of the lengths of both operands. Prediction begins from the node 6. According to the equation (7), overflow occurs on nodes 6, 7 and 9. The data formats of fixed-point partitioning for all nodes are given in Table 1. Notice that truncation error occurs in those nodes because only 16 bits available in machine word and designer must reserve at least one bit for sign (maintaining two's complement representation).

TABLE I.    Fixed-point partitioning of FIR filter's data flow graph from Fig. 6.

| Node | Notation | Node | Notation |
|---|---|---|---|
| $x_n$ | (9/0/7) | 1 | (1/015) |
| $x_{n-1}$ | (9/0/7) | 2 | (1/015) |
| $x_{n-2}$ | (9/0/7) | 3 | (1/015) |
| $x_{n-3}$ | (9/0/7) | 4 | (1/015) |
| $x_{n-4}$ | (9/0/7) | 5 | (1/015) |
| $w_0$ | (8/0/8) | 6 | (1/015), error: $2^{-16} = 0.000015258$ |
| $w_1$ | (8/0/8) | 7 | (1/015), error: $2^{-16} + 2^{-17} = 0.000022888$ |
| $w_2$ | (8/0/8) | 8 | (1/015), error: $2^{-16} + 2^{-17} = 0.000022888$ |
| $w_3$ | (8/0/8) | 9 | (1/015), error: $2^{-16} + 2^{-17} + 2^{-18} = 0.000026702$ |
| $w_4$ | (8/0/8) | | |

This example is quite simple and straightforward but when the size of the filter is large, for example, 128-tap filter, the benefits of overflow prediction is significant.

Equation (7) is not true in case of multiplication where the bit-length of the result is the sum of bit-lengths of operands. Multiplication chains will be explored in the future work.

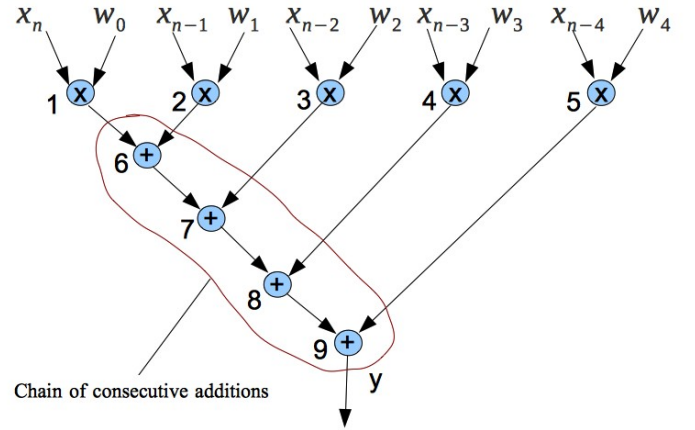

Figure 6. Allocating the chain of consecutive additions.

## III. Experimental results

For exploring the conception of allocating the chains of consecutive additions and equation (7), the special software program was created. Simulation of 10,000 consecutive additions was performed with 8 bit unsigned integer. In Table II below the simulation result are given. The middle column shows the values of *k*, i.e. shows by how much the bit-length of the result increases after each overflow.

TABLE II. 10,000 CONSECUTIVE ADDITIONS.

| Positions (steps) where the overflow occurs | $k$ | Bit-length of the result |
|---|---|---|
| 1 | 1 | 9 |
| 2 | 2 | 10 |
| 4 | 3 | 11 |
| 8 | 4 | 12 |
| 16 | 5 | 13 |
| 32 | 6 | 14 |
| 64 | 7 | 15 |
| 128 | 8 | 16 |
| 257 | 9 | 17 |
| 514 | 10 | 18 |
| 1028 | 11 | 19 |
| 2056 | 12 | 20 |
| 4112 | 13 | 21 |
| 8224 | 14 | 22 |

This conception was implemented in the new generation of SIFOpt software tool [1]. However, the additional work is needed in order to explore all pros and cons of it and in particular, equation (7).

## IV. Conclusion and future work

Allocating the chains of consecutive additions is the powerful predictive technique for designing the fixed-point data paths. It will be explored further in SIFOpt fixed-point data path synthesis tool. Future work will be around exploring consecutive multiplication and improving the presented conception.